\documentclass[aps,showpacs]{revtex4-1}
\begin{document}
\title{ Exact Theorems Concerning CP and CPT Violations in $C=-1$ Entangled State of Pseudoscalar Neutral Mesons
}
\author{Yu Shi}
\email{yushi@fudan.edu.cn}
\affiliation{Department of Physics, Fudan University, Shanghai
200433, China}

\begin{abstract}
Neutral pseudoscalar mesons in an entangled or Einstein-Podolsky-Rosen state are routinely produced in $\phi$ and $B$ factories. Based on the peculiar properties of an entangled state, we present some general exact theorems about  parameters characterizing CP and CPT violations, by using various asymmetries defined for the correlated decays of the two entangled mesons, which are rigorously calculated.
\end{abstract}


Euro. Phys. J. C {\bf 72}, 1907 (2012)

\maketitle

\section{Introduction}

The pair of neutral pseudoscalar mesons produced from a source with total $C=-1$ is in an  entangled or Einstein-Podolsky-Rosen correlated state~\cite{epr}, which cannot be factorized as a direct product of the states of the two mesons~\cite{old,dunietz,bernabeu1,buchanan,dambrosio}.  This fact  has been experimentally confirmed in $K^0\bar{K}^0$ pairs produced in
proton-antiproton annihilation in the CPLEAR detector in
CERN~\cite{cern,cplearrev}, in $K^0\bar{K}^0$ pairs produced at $\phi$ decay
in the KLOE detector in DA$\Phi$NE~\cite{kloe,kloe2,domenico0,domenico}, and in $B_d^0\bar{B}_d^0$ pairs produced at $\Upsilon (4s)$ in the BELLE detector in the KEKB~\cite{belle}. In addition to their values in the perspective of quantum foundations and quantum information~\cite{test,shi}, they have  been in practical use in $\phi$ and $B$ factories such as KLOE, BABAR, BELLE and CLEO,  where neutral pseudoscalar mesons are mostly produced in pairs, and one member of an EPR pair is actually tagged by identifying the other through the product of its decay or interaction with another particle~\cite{kloe2,domenico0,domenico,bigi,cleo}.  A very interesting and important application of the entanglement of meson pairs is the measurement or constraint on the CP and CPT violating parameters
\cite{dunietz,bernabeu,buchanan,dambrosio,domenico0,domenico,kloe2,bernabeu,mav,baji,petrov,soni,kittle,huang}.

In this paper, we present some exact theorems concerning the use of  the $C=-1$ entangled state of pseudoscalar mesons to examine CP and CPT symmetries and extracting parameters characterizing CP and CPT.     The paper is organized as follows. In Sec.~\ref{base}, we review various basis states and the quantum mechanical description of the time-dependent states with  various initial states   of a single neutral pseudoscalar meson. We describe the time evolution of an entangled state in Sec.~\ref{entangle}. We  consider the decays or projections of both mesons into flavor eigenstates in Sec.~\ref{joint}, and give in Sec.~\ref{violation} the method of studying CP and possible CPT violations by using these  events.  Then we consider decays or projections of both mesons into CP eigenstates in Sec.~\ref{cp}, and discuss in Sec~\ref{cp2} the method of studying CP and possible CPT violations by using these events. A summary is made in Sec.~\ref{summary}.

\section{Review of Basis States and Time Evolution of  Single neutral pseudoscalar mesons  \label{base}}

First we give a review of single neutral pseudoscalar mesons in the usual Wigner-Weisskopf approximation, partly to set up definitions and notations, which we find convenient. Some of them follow those in Chou {\it et al.}~\cite{wu}. The relation with some more commonly used definitions and notations will be given explicitly.

A neutral pseudoscalar meson $M^0$ and its antiparticle $\bar{M}^0$ can be described as  eigenstates
of parity $P$ both with eigenvalue $-1$, and of a characteristic flavor with eigenvalues $\pm 1$,  the characteristic flavor being strangeness for $K^0$ and $\bar{K}^0$, beauty for
$B_d^0$ and $\bar{B}_d^0$, charm for $D^0$ and $\bar{D}^0$, and strangeness or beauty (with a minus sign) for   $B_s^0$ and $\bar{B}_s^0$.
As $C|M^0\rangle = -|\bar{M}^0\rangle$ and
$C|\bar{M}^0\rangle = -|M^0\rangle$,  the eigenstates of $CP$ are
\begin{equation}
|M_\pm \rangle =
\frac{1}{\sqrt{2}}(|M^0\rangle \pm  |\bar{M}^0\rangle),
\end{equation}
with eigenvalues $\pm 1$.

Under the Wigner-Weisskopf approximation,
the evolution of an arbitrary state of a pseudoscalar meson $|M(t)\rangle$, as a superposition of $|M^0\rangle$ and $\bar{M}^0\rangle$,  can be described by a Schr\"{o}dinger equation
\begin{equation}
i\frac{\partial}{\partial t}|M(t)\rangle = H |M(t)\rangle,
\end{equation}
where $H$ is the mass matrix with complex elements  $H_{00} \equiv \langle M^0|H|M^0\rangle$,  $H_{0\bar{0}}\equiv\langle M^0|H|\bar{M}^0\rangle $,  $H_{\bar{0}0 } \equiv \langle\bar{ M}^0|H|M^0\rangle$ and $H_{\bar{0}\bar{0}} \equiv \langle\bar{ M}^0|H|\bar{M}^0\rangle$.
CP violation is characterized by a nonzero parameter $\epsilon_M$ defined through
\begin{equation}
\frac{q}{p} \equiv \sqrt{ \frac{H_{\bar{0}0}}{ H_{0\bar{0}}} } \equiv \frac{1-\epsilon_M}{1+\epsilon_M}.
\end{equation}
On the other hand, an indirect CPT violation  would be characterized by a nonzero parameter $\delta_M$ defined as
\begin{equation}
\delta_M \equiv \frac{ H_{\bar{0}\bar{0}} -H_{00}}{ \sqrt{ H_{0\bar{0}}H_{\bar{0}0} } } \neq 0.
\end{equation}
The kind of T violation which is independent of CPT is characterized by whether $|q|/|p|=1$.

The eigenvalues of $H$ are
\begin{eqnarray}
\lambda_S  \equiv m_S -i\Gamma_S/2= H_{00} +  \sqrt{ H_{0\bar{0}}H_{\bar{0}0} } (\sqrt{1+\frac{\delta_M^2}{4}} + \frac{\delta_M}{2}) \\
\lambda_L \equiv  m_L -i\Gamma_L/2
= H_{\bar{0}\bar{0}}-\sqrt{ H_{0\bar{0}}H_{\bar{0}0} }
(\sqrt{1+\frac{\delta_M^2}{4}} + \frac{\delta_M}{2}),
\end{eqnarray}
corresponding, respectively, to its eigenstates
\begin{eqnarray}
|M_S\rangle =
\frac{1}{\sqrt{|p_S|^2+|q_S|^2}}(p_S|M^0\rangle
+q_S|\bar{M}^0\rangle) = \frac{1}{\sqrt{1+|\epsilon_S|^2}}(|M_+\rangle
+\epsilon_S|M_-\rangle), \label{ms1}\\
|M_L\rangle  =
\frac{1}{\sqrt{|p_L|^2+|q_L|^2}}(p_L|M^0\rangle - q_L|\bar{M}^0\rangle) =  \frac{1}{\sqrt{1+|\epsilon_L|^2}}(
\epsilon_L|M_+\rangle+ |M_-\rangle), \label{ml1}
\end{eqnarray}
with
\begin{eqnarray}
x_S \equiv \frac{q_S}{p_S} \equiv \frac{1-\epsilon_S}{1+\epsilon_S}=
\frac{q}{p}(\sqrt{1+\frac{\delta_M^2}{4}}+\frac{\delta_M}{2}),\\
x_L \equiv \frac{q_L}{p_L} \equiv \frac{1-\epsilon_L}{1+\epsilon_L}=
\frac{q}{p}(\sqrt{1+\frac{\delta_M^2}{4}}-\frac{\delta_M}{2}),
\end{eqnarray}
If $\epsilon_S=0$, $|M_S\rangle$   would reduce to $|M_+\rangle$ and $|M_-\rangle$. If $\epsilon_L=0$,  $|M_L\rangle$ would reduce to $|M_-\rangle$.

Many authors use the definitions
\begin{equation}
\epsilon \equiv \frac{1}{2}(\epsilon_S+\epsilon_L),
\end{equation}
\begin{equation}
\delta \equiv \frac{1}{2}(\epsilon_S-\epsilon_L).
\end{equation}
It is straightforward to find
\begin{equation}
\epsilon = \frac{\epsilon_M}{1+\epsilon_M^2+(1-\epsilon_M^2)\sqrt{1+\frac{\delta_M^2}{4}}}
\approx \epsilon_M,
\end{equation}
\begin{equation}
\delta   = -\frac{(1-\epsilon_M^2)\delta_M}{1+\epsilon_M^2+(1-\epsilon_M^2)
\sqrt{1+\frac{\delta_M^2}{4}}}\approx -\frac{\delta_M}{4}.
\end{equation}

We now consider time evolution of a single pseudoscalar neutral meson under the mass Hamiltonian $H$. We summarize the situation by giving the time-dependent state starting with each flavor or CP eigenstate.

Starting as a mass eigenstate $|M_S\rangle$, the state  of a single meson  evolves as
\begin{equation}
|M_S(t) \rangle = e^{-i\lambda_S t} |M_S\rangle.
\end{equation}
Starting as a mass eigenstate $|M_L\rangle$, the state evolves as
\begin{equation}
|M_L(t) \rangle  =  e^{-i\lambda_L t} |M_L\rangle.
\end{equation}

Starting as a flavor eigenstate $|M^0\rangle$, the state evolves as
\begin{equation}
|M^0(t) \rangle = G_{00}(t) |M^0\rangle
+G_{0\bar{0}}(t)|\bar{M}^0\rangle. \label{m0t}
\end{equation}
Starting as a flavor eigenstate  $|\bar{M}_0\rangle$, the state evolves as
\begin{equation}
|\bar{M}^0(t) \rangle=
G_{\bar{0}0}(t) |M^0\rangle
+G_{\bar{0}\bar{0}}(t)|\bar{M}^0\rangle. \label{m0bart}
\end{equation}
In (\ref{m0t}) and (\ref{m0bart}),
\begin{eqnarray}
G_{00}(t) &\equiv& \frac{e^{-i\lambda_S t}+\Omega e^{-i\lambda_L t}}{1+\Omega},\\ G_{0\bar{0}}(t)& \equiv &\frac{x_S(e^{-i\lambda_S t}- e^{-i\lambda_L t})}{1+\Omega}, \\ G_{\bar{0}0}(t)& \equiv &\frac{x_S^{-1}(e^{-i\lambda_S t}- e^{-i\lambda_L t})}{1+\Omega^{-1}},\\
G_{\bar{0}\bar{0}}(t)& \equiv&
\frac{ e^{-i\lambda_S t}+ \Omega^{-1} e^{-i\lambda_L t}}{1+\Omega^{-1}}  .
\end{eqnarray}
where
\begin{equation}
\Omega \equiv \frac{x_S}{x_L} = \frac{q_Sp_L}{p_Sq_L},
\end{equation}

Starting as a CP eigenstate $|M_{+}\rangle$, the state evolves as
\begin{equation}
|M_{+}(t)\rangle = F_{+ +}(t) |M_+\rangle + F_{ +-}(t) |M_-\rangle. \label{mtime}
\end{equation}
Starting as a CP eigenstate $|M_{-}\rangle$, the state evolves as
\begin{equation}
|M_{-}(t)\rangle = F_{- +}(t) |M_+\rangle + F_{- -}(t) |M_-\rangle. \label{mtime2}
\end{equation}
In (\ref{mtime}) and (\ref{mtime2}),
\begin{eqnarray}
F_{++}( t) &=& \frac{1}{2(1+\Omega)}[(1 +x_L^{-1} + x_S + \Omega) e^{-i\lambda_S  t} +(1 -x_L^{-1} - x_S + \Omega) e^{-i \lambda_L  t}],\\
F_{+-}(t) & =& \frac{1 + x_L^{-1} - x_S - \Omega}{2(1+\Omega)}(e^{-i\lambda_S t}- e^{-i \lambda_L t}),\\
F_{-+}( t)& =& \frac{1 - x_L^{-1} + x_S - \Omega}{2(1+\Omega)}(e^{-i\lambda_S t} - e^{-i \lambda_L  t}),\\
F_{--}( t) &=& \frac{1}{2(1+\Omega)}[(1 - x_L^{-1}- x_S + \Omega) e^{-i\lambda_S  t} +(1 + x_L^{-1} + x_S + \Omega) e^{-i \lambda_L  t}].
\end{eqnarray}

\section{Evolution of an entangled state \label{entangle} }

The entangled state of a pair of pseudoscalar mesons produced from a source of $J^{PC}=1^{--}$ is
\begin{eqnarray}
|\Psi_-\rangle & = & \frac{1}{\sqrt{2}}(|M^0\rangle_a|\bar{M}^0\rangle_b
-|\bar{M}^0\rangle_a|M^0\rangle_b)
\\&=& \frac{1}{\sqrt{2}}(|M_-\rangle_a|M_+\rangle_b-
|M_+\rangle_a|M_-\rangle_b).
\end{eqnarray}
Note that not only in the flavor basis, but also in the CP basis, $|\Psi_-\rangle$ is an exact
singlet, no matter whether CP or CPT is  violated.

Now we consider the following situation. Starting as $|\Psi_-\rangle$, the state of the entangled pair evolves and then decay to or produce   certain products at $t_a$ and $t_b$, respectively, which may or may not be equal.  To account for this situation, one first consider
\begin{eqnarray}
|\Psi_-(t_a,t_b)\rangle &=& U_a(t_a) U_b(t_b)|\Psi_-\rangle \\ &=&\frac{1}{\sqrt{2}}(|M^0(t_a)\rangle_a|\bar{M}^0(t_b)\rangle_b
-|\bar{M}^0(t_a)\rangle_a|M^0(t_b)\rangle_b)\\
&=&\frac{1}{\sqrt{2}}(|M_+(t_a)\rangle_a|M_-(t_b)\rangle_b
-|M_-(t_a)\rangle_a|M_+(t_b)\rangle_b).
\end{eqnarray}

Let us refer to these two mesons as  Alice and Bob.
The joint  probability that Alice decays to $|\psi_a\rangle$ at $t_a$ while Bob decays to $|\psi_b\rangle$ at $t_b$ was calculated as
\begin{equation}
I(\psi_a,t_a;\psi_b,t_b) = |\langle \psi_a,\psi_b |\Psi(t_a,t_b)\rangle|^2,
\end{equation}
where
\begin{equation}
|\langle\psi_a,\psi_b|\Psi_-(t_a,t_b)\rangle
=\frac{1}{\sqrt{2}}(\langle\psi_a|M_+(t_a)\rangle_a \langle \psi_b |M_-(t_b)\rangle_b
- \langle\psi_a|M_-(t_a)\rangle_a \langle \psi_b |M_+(t_b)\rangle_b).
\end{equation}

Various situations are discussed in the following.

\section{Joint probabilities of the decays into  flavor eigenstates \label{joint} }

Suppose from the entangled state, Alice and Bob each decays or transits to a two-valued flavor eigenstate,  as in CPLEAR experiment, where the strangeness values of the two kaons  were measured through their interactions with bound nucleons~\cite{cern,cplearrev},
and in BELLE experiment, where the beauties of the two $B_d$ mesons  were  measured by their semileptonic decays~\cite{belle}. We generally denote the flavor eigenstates as $|l^+\rangle$ of eigenvalue $+1$ and  $|l^-\rangle$ of eigenvalue $-1$.  Examples for  $|l^+\rangle$ include the semileptonic decay product $M^-\bar{l}\nu$, as well as $D^-D_S^+$, $D^-K^+$, $\pi^-D_S^+$, $\pi^-K^+$ from $M^0=B^0$, and  $D_S^-\pi^+$,   $D_S^-D^+$,  $K^-\pi^+$,  $K^-D^+$ from $M^0=B_S^0$.   Examples for  $|l^-\rangle$ include the semileptonic decay product $M^+ l\bar{\nu}$, as well as $D^+D_S^-$, $D^+K^-$, $\pi^+D_S^-$, $\pi^+K^-$ from $\bar{M}^0=B^0$, and  $D_S^+\pi^-$,   $D_S^+D^-$,  $K^+\pi^-$,  $K^+D^-$ from $M^0=B_S^0$~\cite{wu}. In the CPLEAR experiment on kaons~\cite{cern}, $|l^+\rangle$ and $|l^-\rangle$ are  products produced via interaction with bound nucleons.

For   $|l^+\rangle$,  the  amplitude from  $|M^0\rangle$  is
\begin{equation}
\langle l^+|M^0\rangle\equiv R^+ \neq 0
\end{equation}
while the amplitude from  $\bar{M}^0$ is
\begin{equation}
\langle l^+|\bar{M}^0\rangle = 0.
\end{equation}
For a flavor eigenstate $|l^-\rangle$, the amplitude from  $\bar{M}^0$ is
\begin{equation}
\langle l^-|\bar{M}^0\rangle\equiv R^- \neq 0
\end{equation}
while  the  amplitude from  $M^0\rangle$  is
\begin{equation}
\langle l^-|M^0\rangle = 0.
\end{equation}

One can calculate the joint amplitude of such a pair of decays in which Alice and Bob decays $| l_a^x\rangle $  at $t_a$ while Bob decays to $|l_b^y\rangle $  at $t_b$,  where $x$ and $y$ each represents $\pm 1$,
\begin{equation}
\langle l_a^x,l_b^y|\Psi_-(t_a,t_b)\rangle =  \frac{1}{\sqrt{2}}(\langle l_a^x|M^0(t_a)\rangle_a \langle l_b^y|\bar{M}^0(t_b)\rangle_b
-\langle l_a^x|\bar{M}^0(t_a)\rangle_a  \langle l_b^y|M^0(t_b)\rangle_b).
\end{equation}

From (\ref{m0t}) and (\ref{m0bart}), one can obtain
\begin{eqnarray}
\langle l_{\alpha}^+|M^0(t_\alpha)\rangle_\alpha &= &G_{00}(t_\alpha)R^+_\alpha ,\\
\langle l_{\alpha}^-|M^0(t_\alpha)\rangle_\alpha &= &G_{0\bar{0}}(t_\alpha)R^-_\alpha,\\
\langle l_{\alpha}^+|\bar{M}^0(t_\alpha)\rangle_\alpha &= &G_{\bar{0}0}(t_\alpha)R^+_\alpha,\\
\langle l_{\alpha}^-|\bar{M}^0(t_\alpha)\rangle_\alpha &= &G_{\bar{0}\bar{0}}(t_\alpha)R^-_\alpha,
\end{eqnarray}
where $\alpha =a,b$.

Thus we obtain
\begin{equation}
\begin{array}{rcl}
I(l^+_a, t_a; l^+_b, t_b)&=& \frac{|R^+_aR^+_b|^2|x_L|^{-2}}{2|1+\Omega|^2} e^{-(\Gamma_S+\Gamma_L)t_a}
[e^{-\Gamma_S\Delta t}+ e^{-\Gamma_L\Delta t} -2e^{\frac{\Gamma_S+\Gamma_L}{2}\Delta t} \cos (\Delta m \Delta t)], \\
I(l^+_a, t_a; l^{-}_b, t_b)&=&  \frac{|R^+_aR^b_-|^2}{2|1+\Omega|^2}e^{-(\Gamma_S+\Gamma_L)t_a}
[|\Omega|^{2}e^{-\Gamma_S\Delta t} + e^{-\Gamma_L\Delta t} +2|\Omega|e^{\frac{\Gamma_S+\Gamma_L}{2} \Delta t} \cos (\Delta m \Delta t +\phi_{\Omega})],\\
I(l^-_a, t_a; l^+_b, t_b) &=&
\frac{|R_a^-R^+_b|^2}{2|1+\Omega|^2}e^{-(\Gamma_S+\Gamma_L)t_a}
[e^{-\Gamma_S\Delta t}   +|\Omega|^2 e^{-\Gamma_L\Delta t} +2|\Omega|e^{\frac{\Gamma_S+\Gamma_L}{2} \Delta t}
 \cos (\Delta m \Delta t -\phi_{\Omega})],   \\
I(l^-_a, t_a; l^-_b, t_b)& =&  \frac{|R_a^-R_b^-|^2|x_S|^2}{2|1+\Omega|^2}e^{-(\Gamma_S+\Gamma_L)t_a}  [e^{-\Gamma_S\Delta t} + e^{-\Gamma_L\Delta t} -2e^{\frac{\Gamma_S+\Gamma_L}{2}\Delta t} \cos (\Delta m \Delta t)],
\end{array} \label{il}
\end{equation}
where $\Delta m \equiv m_L-m_S$, $\Delta t \equiv t_b-t_a $, $\Omega = |\Omega|e^{i\phi_{\Omega}}$.

In experiments, it is more convenient to use the integrated rate
\begin{equation}
I'(l_a^x, l_b^y, \Delta t) = \int_0^\infty I(l_a^x, t_a; l_b^y, t_a+\Delta t) dt_a,
\end{equation}
which is simply given by $I(l^x_a, t_a; l^y_b, t_a+\Delta t)$ as in (\ref{il}),  with $e^{-(\Gamma_S+\Gamma_L)t_a}$ replaced as $1/(\Gamma_S+\Gamma_L)$.

\section{Studying CP and CPT violations using the joint probabilities of decays in flavor basis  \label{violation}}

We focus on the case $|l_a^+\rangle = |l_b^+\rangle$ and  $|l_a^-\rangle = |l_b^-\rangle$, hence $R^+_a=R^+_b=R^+$, $R_a^-=R_b^-=R^-$.

The above four joint probabilities can form some asymmetries between the decays of the entangled mesons.

First consider
\begin{equation}
A(++,--, t_a, t_b) \equiv \frac{I[l^+_a,t_a; l^+_b,t_b]-I[l^-_a,t_a; l^-_b,t_b]}{I[l^+_a,t_a; l^+_b,t_b]+ I[l^-_a,t_a; l^-_b,t_b]},
\end{equation}
and the asymmetry of the corresponding integrated rates
\begin{equation}
A'(++,--, \Delta t) \equiv \frac{I'[l^+_a, l^+_b, \Delta t]-I'[l^-_a, l^-_b, \Delta t]}
{I'[l^+_a, l^+_b, \Delta t]+I'[l^-_a, l^-_b, \Delta t]},
\end{equation}
which are only dependent on $\Delta t \equiv t_b-t_a$.

They are evaluated to be
\begin{equation}
A(++,--,t_a,t_b) = A'(++,--, \Delta t) = \frac{{|R^+|}^4|x_L|^{-1}- {|R^-|}^4|x_S|}{{|R^+|}^4|x_L|^{-1}+ {|R^-|}^4|x_S|}, \label{a10}
\end{equation}
which is independent of $t_a$,  $t_b$ or  $\Delta t$.

If CP is conserved, then $R^+=R^-$, $p/q=1$, thus $x_S=x_L^{-1}$, consequently  $A(++,--,t_a,t_b) = A'(++,--, \Delta t) = 0$.  Therefore we have the following theorem.

{\bf Theorem 1}
   {\em If the equal-flavor asymmetry $A(++,--,t_a,t_b) \neq 0$, or if $A'(++,--, \Delta t) \neq 0$,  then CP is  violated, independent of whether CPT is conserved.}

On the other hand, if $CPT$ is conserved, then $\delta_M=0$, thus $x_S=x_L=q/p$, consequently  we have the following result.

{\bf Theorem 2} {\em If CPT is conserved, the equal-flavor asymmetry is $A(++,--,t_a,t_b)$ or $A'(++,--, \Delta t)$ is given by      }

\begin{equation}
A(++,--,t_a,t_b) = A'(++,--, \Delta t) = \frac{{|R^+|}^4|p|^2- {|R^-|}^4|q|^2 }{{|R^+|}^4||p|^2 + {|R^-|}^4|q|^2}. \label{a1}
\end{equation}
A deviation from (\ref{a1}) indicates  CPT violation, which, however, can be tested by the following simpler method.

Now we consider the unequal-flavor asymmetry defined by
\begin{equation}
A(+-,-+, t_a, t_b) \equiv \frac{I[l^+_a,t_a; l^-_b,t_b]-I[l^-_a,t_a; l^+_b,t_b]}{I[l^+_a,t_a; l^-_b,t_b]+ I[l^-_a,t_a; l^+_b,t_b]},
\end{equation}
and the asymmetry of the corresponding integrated rates
\begin{equation}
A'(+-,-+, \Delta t) \equiv \frac{I'[l^+_a, l^-_b,\Delta t]-I'[l^-_a, l^+_b,\Delta t]}{I[l^+_a, l^-_b,\Delta t]+ I[l^-_a, l^+_b,\Delta t]},
\end{equation}
which are evaluated to be
\begin{eqnarray}
&A(+-,-+, t_a, t_b) = A'(+-,-+, \Delta t) \nonumber \\
&=  \frac{(1-|\Omega|^2) ( e^{-\Gamma_L\Delta t} -e^{-\Gamma_S\Delta t})+ 4 |\Omega| e^{\frac{\Gamma_S+\Gamma_L}{2}\Delta t} \sin (\Delta m \Delta t) \sin \phi_{\Omega}}{(1+|\Omega|^2) ( e^{-\Gamma_L\Delta t} -e^{-\Gamma_S\Delta t})+ 4 |\Omega| e^{\frac{\Gamma_S+\Gamma_L}{2}\Delta t} \cos (\Delta m \Delta t) \cos \phi_{\Omega}},
\end{eqnarray}
which always vanishes if $\Delta t =0$. For  $\Delta t \neq 0$,
if CPT is conserved, then $\Omega =1$, i.e. $|\Omega|=1$ and $\sin\phi_\Omega=0$, consequently  $A(+-,-+, t_a, t_b) = A'(+-,-+, \Delta t) =0$.  Therefore we have the following theorem.

{\bf Theorem 3} {\em Consider the two correlated decays at different times. If the unequal-flavor asymmetry
$A(+-,-+, t_a, t_b)\neq 0$, or if  $A'(+-,-+,\Delta t)\neq 0$,  then CPT is violated.}

Moreover, using the four instantaneous rates at various times or using the four integrated rates for various time differences, one can determine all the parameters $|R^+|$, $|R^-|$, $|x_S|$, $|x_L|$, $|\Omega|$, $\phi_{\Omega}$, $\Gamma_S$, $\Gamma_L$, $\Delta m$, which are all rephase-invariant.
One can also simply use the integrates rates with various values of $\Delta t$. One can also use the instantaneous rates while avoid using the precise value of $t_a$, by considering the ratio
$
I(l^+_a, t_a; l^+_b, t_b):
I(l^+_a, t_a; l^{-}_b, t_b):
I(l^-_a, t_a; l^+_b, t_b) :
I(l^-_a, t_a; l^-_b, t_b)$, where the common factor $e^{-(\Gamma_S+\Gamma_L)t_a}$ is canceled.  Therefore the instantaneous rates with the same $\Delta t$ but different $t_a$ can be used altogether.

\section{Joint probabilities of decays into   CP eigenstates \label{cp} }

Now we consider the situation that the decay products of Alice and Bob  are both CP eigenstates, as in  KLOE experiment, where the probability of both kaons decaying to  $\pi^+\pi^-$ was obtained up to a proportional factor~\cite{kloe,domenico}.  We generally denote the CP eigenstates as $|h^+\rangle$ of eigenvalue $+1$ and  $|h^-\rangle$ of eigenvalue $-1$.  Examples for  $|h^+\rangle$ include    $\pi^+\pi^-$, $\pi^0\pi^0$, etc.    Examples for  $|h^-\rangle$ include  $\pi^0\pi^0\pi^0$, etc.

For $|h^+\rangle$, the amplitude from  $|M_+\rangle$ is
\begin{equation}
\langle h^+|M_+\rangle\equiv Q^+ \neq 0,
\end{equation}
while the amplitude from  $|M_-\rangle$ is
\begin{equation}
\langle h^+|M_-\rangle = 0.
\end{equation}
For   $|h^-\rangle$,  the amplitude from  $|M_-\rangle$ is
\begin{equation}
\langle h^-|M_-\rangle\equiv Q^- \neq 0
\end{equation}
while the amplitude from  $|M_+\rangle$ is
\begin{equation}
\langle h^-|M_+\rangle = 0.
\end{equation}

$Q^{+}$ and $Q^-$ are not  directly measurable quantities, as $M_S$ and $M_L$, rather than $M_+$ and $M_-$, are physical. However, from (\ref{ms1}) and (\ref{ml1}), we have
\begin{eqnarray}
Q^+&=&\sqrt{1+|\epsilon_S|^2} \langle h^+|M_S\rangle, \\
Q^-&=&\sqrt{1+|\epsilon_L|^2} \langle h^-|M_L\rangle.
\end{eqnarray}
It is not difficult to see that although $Q^{\pm}$ are not rephase-invariant,  $|Q^{\pm}|^2$ are.

Suppose Alice and Bob decay to CP eigenstates $h_a^x$ and $h_b^y$ respectively, where $x$ and $y$ each represents $\pm 1$. Then
\begin{equation}
\langle h_a^x,h_b^y|\Psi_-(t_a,t_b)\rangle =  \frac{1}{\sqrt{2}}(\langle h_a^x|M_+(t_a)\rangle_a \langle h_b^y|M_-(t_b)\rangle_b
-\langle h_a^x|M_-(t_a)\rangle_a  \langle h_b^y|M_+(t_b)\rangle_b).
\end{equation}

From (\ref{mtime}) and (\ref{mtime2}), one can obtain
\begin{eqnarray}
\langle h_{\alpha}^+|M_+(t_\alpha)\rangle_\alpha &= &F_{++}(t_\alpha)Q^+_\alpha,\\
\langle h_{\alpha}^-|M_+(t_\alpha)\rangle_\alpha &= &F_{+-}(t_\alpha)Q^-_\alpha,\\
\langle h_{\alpha}^+|M_-(t_\alpha)\rangle_\alpha &= &F_{-+}(t_\alpha)Q^+_\alpha,\\
\langle h_{\alpha}^-|M_-(t_\alpha)\rangle_\alpha &= &F_{--}(t_\alpha)Q^-_\alpha.
\end{eqnarray}

Therefore,

\begin{equation}
\begin{array}{rcl}
I[h^+_a, t_a; h^+_b, t_b] & = &
\frac{|Q_{+a}|^2 |Q_{+b}|^2 |(1 - x_L^{-1})(1 + x_S)|^2}{8|1+\Omega|^2}   e^{-(\Gamma_S+\Gamma_L)t_a}\\
&&\times [e^{-\Gamma_S (\Delta t)} + e^{-\Gamma_L(\Delta t)} - 2 e^{-\frac{1}{2}(\Gamma_S+\Gamma_L)\Delta t} \cos(\Delta m \Delta t)], \\
I[h^+_a, t_a; h^-_b, t_b] & = &
\frac{|Q_{+a}|^2 |Q_{-b}|^2|(1 +x_L^{-1})(1+x_S)|^2}{8|1+\Omega|^2} e^{-(\Gamma_S+\Gamma_L)t_a} \\
&&\times  [|W|^{2}e^{-\Gamma_S (\Delta t)} +e^{-\Gamma_L(\Delta t)} + 2 |W| e^{-\frac{1}{2}(\Gamma_S+\Gamma_L)\Delta t} \cos(\Delta m \Delta t+\phi_W)],  \\
I[h^-_a, t_a; h^+_b, t_b]  & = &
\frac{ |Q_{-a}|^2 |Q_{+b}|^2 |(1+x_L^{-1})(1+x_S)|^2}{8|1+\Omega|^2}  e^{-(\Gamma_S+\Gamma_L)t_a}\\
&&\times   [e^{-\Gamma_S (\Delta t)}  +  |W|^2 e^{-\Gamma_L(\Delta t)}+ 2 |W| e^{-\frac{1}{2}(\Gamma_S+\Gamma_L)\Delta t} \cos(\Delta m \Delta t-\phi_W)],   \\
I[h^-_a, t_a; h^-_b, t_b] & = &
\frac{|Q_{-a}|^2  |Q_{-b}|^2 |(1+x_L^{-1})(1-x_S)|^2}{8|1+\Omega|^2} e^{-(\Gamma_S+\Gamma_L)t_a}\\
&&\times  [e^{-\Gamma_S (\Delta t)} + e^{-\Gamma_L(\Delta t)} - 2 e^{-\frac{1}{2}(\Gamma_S+\Gamma_L)\Delta t} \cos(\Delta m \Delta t)],
\end{array} \label{i2}
\end{equation}
where
\begin{equation}
W \equiv \frac{(1-x_L^{-1})(1-x_S)}{(1+x_L^{-1})(1+x_S)}.
\end{equation}

One can also consider the integrated rate
\begin{equation}
I'(h_a^x, h_b^y, \Delta t) = \int_0^\infty I(h^x_a, t_a; h^y_b, t_a+\Delta t) dt_a,
\end{equation}
which is simply given by $I(h^x_a, t_a; h^y_b, t_a+\Delta t)$ as in (\ref{i2}),  with $e^{-(\Gamma_S+\Gamma_L)t_a}$ replaced as $1/(\Gamma_S+\Gamma_L)$.

\section{ Studying CP and CPT violations using the joint probabilities of decays in CP basis   \label{cp2}}

First,  if both CP and CPT are conserved, then  $x_S=x_L^{-1} =1$, consequently   $I[h^+_a, t_a; h^+_b, t_b]=I[h^-_a, t_a; h^-_b, t_b]=I'[h^+_a, h^+_b, \Delta t]=[h^-_a,  h^-_b,\Delta t]=0$ even if $\Delta t \neq 0$. For $\Delta t=0$, we always have $I[h^+_a, t_a; h^+_b, t_a]=[h^-_a, t_a; h^-_b, t_a]=I'[h^+_a, h^+_b, 0]=[h^-_a,  h^-_b,0]=0$   no matter whether CP or CPT is violated. Therefore we have the following theorem.

{\bf Theorem 4} {\em
Any nonzero value of $I[h^+_a, t_a; h^+_b, t_b]$ or $I[h^-_a, t_a; h^-_b, t_b]$ or $I'[h^+_a, h^+_b, \Delta t]$ or $I[h^-_a,  h^-_b,\Delta t]$ for  $\Delta t\neq 0$ indicates the violation of CP or CPT or both.}

KLOE experimental data~\cite{domenico0,domenico} indeed fit well $$I(h^+_a, t_a; h^+_b, t_b) \propto e^{-\Gamma_S (\Delta t)} + e^{-\Gamma_L(\Delta t)} - 2 e^{-\frac{1}{2}(\Gamma_S+\Gamma_L)\Delta t} \cos(\Delta m \Delta t),$$
which should be regarded as an indication of CP or CPT or both.

Now let us consider the asymmetries, focusing on the case  $|h_a^+\rangle = |h_b^+\rangle$ and  $|h_a^-\rangle = |h_b^-\rangle$, hence $Q^+_a=Q^+_b=Q^+$, $Q_a^-=Q_b^-=Q^-$.

First consider the asymmetry
\begin{equation}
B(++,--,t_a,t_b) \equiv \frac{I[h^+_a,t_a; h^+_b,t_b]-I[h^-_a,t_a; h^-_b,t_b]}{I[h^+_a,t_a; h^+_b,t_b]+ I[h^-_a,t_a; h^-_b,t_b]}.
\end{equation}
and the corresponding  one for the integrated rates
\begin{equation}
B'(++,--, \Delta t) \equiv \frac{I'[h^+_a, h^+_b, \Delta t]-I'[h^-_a, h^-_b, \Delta t]}
{I'[h^+_a, h^+_b, \Delta t]+I'[h^-_a, h^-_b, \Delta t]}.
\end{equation}
which are evaluated to be
\begin{eqnarray}
&B(++,--,t_a,t_b) = B'(++,--, \Delta t) \nonumber \\
& =  \frac{{|Q^+|}^4|(1 - x_L^{-1})(1 + x_S)|^2 - {|Q^-|}^4 |(1 +x_L^{-1})(1 - x_S)|^2 }{ {|Q^+|}^4|(1 - x_L^{-1})(1 + x_S)|^2 + {|Q^-|}^4 |(1 +x_L^{-1})(1 - x_S)|^2   }, \end{eqnarray}
which are independent of $t_a$, $t_b$ or $\Delta t$.

If CP is conserved, then $p/q=1$, thus $x_S=1/x_L=\sqrt{1+\delta_M^2/4}-\delta_M/2$, thus
\begin{eqnarray}
&B(++,--,t_a,t_b) = B'(++,--, \Delta t) \nonumber \\
& =  \frac{{|Q^+|}^4 - {|Q^-|}^4 }{ {|Q^+|}^4 + {|Q^-|}^4   }, \label{bab}  \end{eqnarray}
which is independent of $t_a$, $t_b$ or $\Delta t$.

On the other hand, if CPT is conserved, then $\delta_M=0$, and thus $x_L=x_S=q/p$, consequently
\begin{eqnarray}
&B(++,--,t_a,t_b) = B'(++,--, \Delta t) \nonumber \\
& =  \frac{{|Q^+|}^4 + {|Q^-|}^4  }{ {|Q^+|}^4 - {|Q^-|}^4    }, \label{bab2} \end{eqnarray}
which is independent of $t_a$, $t_b$ or $\Delta t$, and is in fact the inverse of (\ref{bab}). Note that if both CPT and CP are conserved, then it is not meaningful to construct this asymmetry, as the two relevant joint probabilities are then both zero.

Therefore, while the violation of  CP or CPT or both is indicated by the nonzero value of $I[h^+_a, t_a; h^+_b, t_b]$ or $I[h^-_a, t_a; h^-_b, t_b]$ or $I'[h^+_a, h^+_b, \Delta t]$ or $I[h^-_a,  h^-_b,\Delta t]$, we can further examine the equal-CP asymmetry $B(++,--,t_a,t_b)$ or $ B'(+-,-+, \Delta t)$, and have the following theorem, which recognize which of CP and CPT is violated.

{\bf Theorem 5} {\em Suppose both $I[h^+_a, t_a; h^+_b, t_b]$ and $I[h^-_a, t_a; h^-_b, t_b]$ are nonzero. If  the equal-CP asymmetry  asymmetry $B(++,--,t_a,t_b)$ or $ B'(++,--, \Delta t)$ is given by $\frac{{|Q^+|}^4 - {|Q^-|}^4 }{ {|Q^+|}^4 + {|Q^-|}^4   }$, then CP is conserved but CPT is violated. If it is given by  $\frac{{|Q^+|}^4 + {|Q^-|}^4  }{ {|Q^+|}^4 - {|Q^-|}^4    }$, then CPT is conserved but CP is violated. If it is not given by neither $\frac{{|Q^+|}^4 - {|Q^-|}^4 }{ {|Q^+|}^4 + {|Q^-|}^4   }$   nor $\frac{{|Q^+|}^4 + {|Q^-|}^4  }{ {|Q^+|}^4 - {|Q^-|}^4    }$, then both CP and CPT are violated. }

Now we consider unequal-CP asymmetry
\begin{equation}
B(+-,-+,t_a,t_b) \equiv \frac{I[h^+_a,t_a; h^-_b,t_b]-I[h^-_a,t_a; h^+_b,t_b]}{I[h^+_a,t_a; h^-_b,t_b]+ I[h^-_a,t_a; h^+_b,t_b]}.
\end{equation}
and the corresponding  one for the integrated rates
\begin{equation}
B'(+-,-+, \Delta t) \equiv \frac{I'[h^+_a, h^-_b,\Delta t]-I'[h^-_a, h^+_b,\Delta t]}{I[h^+_a, h^-_b,\Delta t]+ I[h^-_a, h^+_b,\Delta t]},
\end{equation}
which are  evaluated to be
\begin{eqnarray}
&B(+-,-+,t_a,t_b) = B'(+-,-+, \Delta t)\nonumber \\
&= \frac{(1-|W|^2) ( e^{-\Gamma_L\Delta t} -e^{-\Gamma_S\Delta t})+ 4 |W| e^{\frac{\Gamma_S+\Gamma_L}{2}\Delta t} \sin (\Delta m \Delta t) \sin \phi_{W}}{(1+|W|^2) ( e^{-\Gamma_L\Delta t} +e^{-\Gamma_S\Delta t})+ 4 |W| e^{\frac{\Gamma_S+\Gamma_L}{2}\Delta t} \cos (\Delta m \Delta t) \cos \phi_{W}}.
\end{eqnarray}

If both  CPT and CP are conserved, then $W =0$,
\begin{equation}
B(+-,-+,t_a,t_b) = B'(+-,-+, \Delta t)= \frac{e^{-\Gamma_L\Delta t} -e^{-\Gamma_S\Delta t}}{ e^{-\Gamma_L\Delta t} +e^{-\Gamma_S\Delta t}}, \label{as}
\end{equation}
In particular, when $\Delta t=0$, the unequal-CP asymmetry is $0$. Hence we have the following theorem.

{\bf Theorem 6} {\em If the unequal-CP asymmetry $B(+-,-+,t_a,t_b)$ or $B'(+-,-+, \Delta t)$ deviates  from $\frac{e^{-\Gamma_L\Delta t} -e^{-\Gamma_S\Delta t}}{ e^{-\Gamma_L\Delta t} +e^{-\Gamma_S\Delta t}}$, then  CPT or CP or both are violated. In particular, if the equal time unequal-CP asymmetry $B(+-,-+,t_a,t_a)$ or $B'(+-,-+, 0)$ is nonzero, then  CPT or CP or both are violated.}

Finally, of course, using the four instantaneous rates or the four integrated rates for various values of $\Delta t$, one can determine the parameters $|Q^+|$, $|Q^-|$, $|1\pm x_S|$, $|1\pm x_L|$, $|W|$, $\phi_{W}$, $\Gamma_S$, $\Gamma_L$, $\Delta m$.

\section{summary \label{summary} }

In this paper, we have considered the correlated decays of a $C=-1$ entangled state of two pseudoscalar mesons, which are now routinely produced in $\phi$ and $B$ factories. We rigorously calculated the joint probabilities and corresponding integrated rates of the decays both into flavor eigenstates, as well as those of the decays both into CP eigenstates. Measurement of these joint probabilities for various values of time differences of the two decays can lead to the determination of various parameters, including those characterizing CP and CPT violations.

More interesting results are some rigorously formulated simple methods of deciding whether CP and CPT are violated by using the asymmetries of the entangled decays of the two mesons. The asymmetry between instantaneous probabilities (with unprimed notation) is equal to the corresponding one  between integrated rates (with primed notation).

If the equal-flavor asymmetry $A(++,--,t_a,t_b)$ or $A'(++,--, \Delta t) $, which is found to be independent of times,   is nonzero,  then CP is  violated, independent of whether CPT is conserved.  If the unequal-flavor asymmetry $A(+-,-+, t_a, t_b)$ or  $A'(+-,-+,\Delta t)$ is nonzero even if $\Delta t \neq 0$,  then CPT is violated.

Any nonzero value of a joint probability of equal-CP decays at different times, i.e. $I[h^+_a, t_a; h^+_b, t_b]$ or $I[h^-_a, t_a; h^-_b, t_b]$ or $I'[h^+_a, h^+_b, \Delta t]$ or $I[h^-_a,  h^-_b,\Delta t]$   for  $\Delta t \neq 0$, indicates the violation of CP or CPT or both. Hence we emphasize that the nonzero $I[h^+_a, t_a; h^+_b, t_b]$   observed in KLOE only indicates the violation of CP or CPT or both, but cannot decide which of them are actually violated.

Furthermore, if the the equal-CP asymmetry  asymmetry $B(++,--,t_a,t_b)$ or $ B'(+-,-+, \Delta t)$ is given by $\frac{{|Q^+|}^4 - {|Q^-|}^4 }{ {|Q^+|}^4 + {|Q^-|}^4   }$, then CP is conserved but CPT is violated. If it is given by $\frac{{|Q^+|}^4 + {|Q^-|}^4  }{ {|Q^+|}^4 - {|Q^-|}^4    }$, then CPT is conserved but CP is violated. If it satisfies neither $\frac{{|Q^+|}^4 - {|Q^-|}^4 }{ {|Q^+|}^4 + {|Q^-|}^4   }$ nor $\frac{{|Q^+|}^4 + {|Q^-|}^4  }{ {|Q^+|}^4 - {|Q^-|}^4    }$, then both CP and CPT are violated.

If the unequal-CP asymmetry $B(+-,-+,t_a,t_b)$ or $B'(+-,-+, \Delta t)$ deviates  from  $\frac{e^{-\Gamma_L\Delta t} -e^{-\Gamma_S\Delta t}}{ e^{-\Gamma_L\Delta t} +e^{-\Gamma_S\Delta t}}$, then  CPT or CP or both are violated. In particular, if the equal time unequal-CP asymmetry $B(+-,-+,t_a,t_a)$ or $B'(+-,-+, 0)$ is nonzero, then  CPT or CP or both are violated.

We believe that some convenient ways of testing CP and CPT symmetries have been presented here for the use of entangled mesons in $\phi$ and B factories.

\vspace{1cm}

I am very grateful to Y. L. Wu for useful discussions.
This work was supported by the National Science Foundation of China (Grant No. 10875028).

\end{document}